\documentclass[aps,prl,showpacs,twocolumn,groupedaddress,amssymb]{revtex4}
\usepackage{graphicx}% Include figure files
\usepackage{dcolumn}% Align table columns on decimal point
\usepackage{bm}% bold math

\begin{document}

\title{THz light amplification by an visible light laser in the presence of the  plasma density gradient}% \\

\author{S. Son}
\affiliation{169 Snowden Lane, Princeton, NJ, 08540}
%\author{Sung Joon Moon}
%\affiliation{PACM, Princeton University, Princeton, NJ 08544}
%\author{J.~Y. Park}
%\affiliation{Los Alamos National Laboratory}
%\date{\today}% It is always \today, today,
             %  but any date may be explicitly specified

\begin{abstract}
A new mechanism for 
the THz light  amplification is identified in
a non-resonant Raman scattering between the THz light and a  visible light lasers. The non-resonant scattering normally does not exchange the energy between E \&M fields, but 
% but  a pre-exciting Langmuir wave, the author shows, enables the energy transfer from the laser to the x-ray light. % which would not be possible otherwise.
the presence of the plasma density gradient
 creates an condition in which a visible light laser could transfer its energy into  a tera-hertz (THz) light via the laser-plasma interaction. 
 %The THz light amplification in the presence of the plasma  density gradient is identified, 
%where 
%The density perturbation from  the ponderomotive interaction could  have 
% the desirable phase for transferring the energy  from the laser to the THz light. 
%is excited via  the ponderomotive interaction between an intense laser and the THz light. The strong and lasting density gradient is provide by  a  Langmuir wave whose phase velocity is the same with the group velocity of the THz light.  
The gain per length could reach 100 (1000) per centimeter for the THz light (far infra-red light) amplification. 
%The non-linear convection for the THz light generation from  two-colored lasers are analyzed, based on the forward Raman scattering. 
%The energy transfer from the lasers to the THz light can be efficient. The possible peak intensity of the generated THz light is estimated and the optimal duration time is estimated.   
%A new scheme for soft x-ray lasers  is proposed. The backward Haman scattering between an intense visible-light laser and a relativistic electron beam results in soft x-ray light via the Doppler shift. One of the most intense soft x-ray light sources is contemplated.  
\end{abstract}


\maketitle

Even though the great advances have been made in the THz light technologies~\cite{Tilborg,Zheng,Reimann, gyrotron, gyrotron2, gyrotron3,magnetron, qlaser, qlaser3, freelaser, freelaser2, colson, songamma,Gallardo,sonttera}, 
current THz light sources are
still not intense  enough for many applications.
As it is  termed  as the ``THz Gap''~\cite{booske},
the current technologies cannot reach the THz light   intensity, practically desirable or theoretically possible.
% that 
%the current technologies cannot reach the THz light   intensity, practically desirable or theoretically possible.  
  %to the theoretical limit 
%is referred to as the ``THz Gap''~[1].  
If a comparison is made between the progresses in the visible light laser and the THz light technology,  one obvious missing  ingredient is an amplifier. %wherein a small signal gets amplified to an  intense one.
A strong THz amplifier, if exists, could lead to a broad commercialization of the THz light in  various applications. 
In this paper, one possible THz light amplifier is proposed based on the non-resonant Raman scattering between the THz light and an intense visible-light laser.

The Raman scattering is the well-known non-linear laser-plasma interaction~\cite{Fisch, malkin1, sonbackward, BBRS, BBRS2, BBRS3}.   
 The ponderomotive interaction between two lasers excite a Langmuir wave,  transferring the energy from the higher frequency laser to the lower frequency laser. 
A tempting  question would be whether a visible-light laser can amplify a THz light via  a similar mechanism.
%could  transfer the energy from a visible-light laser to a THz light. % which could be an natural THz light amplifier.  
%Of course, the immediate answer is that it is impossible. 
Unfortunately, 
 because of the frequency difference
between a visible light laser and a THz light, 
the beating pondermotive interaction is a non-resonant interaction and   
the non-resonantly excited density perturbation does not have a phase, suitable for the energy transfer between the visible light laser and a THz light.

In this paper, considering  a situation where an intense visible light laser and a THz light are propagating each other in the presence of the strong plasma density gradient, 
the author shows that the ponderomotive density perturbation could have a phase suitable  for the amplification even if it is a non-resonant scattering. 
% This phase mismatch can be utilized to the effect of transferring energy from the visible light laser to the THz light via the beating current. 
% the author contemplate utilizing the density gradient of the plasma in order to achieve the off-cycle phase of the density perturbation to the ponderomov
%the desired phase-lock of the density perturation in the Raman scattering to the benefit of the THz light amplification. 
In most cases, the amplification from the identified process 
is very small  and the time average of the time oscillating density gradient tends to be even smaller along the THz light propagation path. 
However, the author considers a particular  case when 
a  lasting density gradient is provided by  a co-propagating Langmuir wave whose phase velocity is the same with the group velocity of the THz light. 
In this situation, the oscillation will be suppressed so that  a strong THz light amplification is possible. 
%It is for the first time that such an amplifying mechanism is discussed. 
The new scheme proposed in this paper can achieve  
   the gain-per-length as high as  100 (1000) per centimeter for the THz light (far infra-red light) amplification. 
% While quantum cascade laser is an amplifier, it only works in a low-intensity laser. 
%If an intense amplifier is available, many obstacles for the THz applications could be overcome. One such amplifier is proposed in this paper based on the backward Raman scattering. 

Let us consider a THz light with the frequency $\omega_T$   co-propagating or counter-propagating with an intense laser ($\omega_1$) in the z-direction. For simplicity, the THz light and the laser is assumed to be linearly polarized.   Furthermore, let us assume that 
there exists an intense plasma wave $n_L(z-v_Tt)$, which has the same phase velocity with the group velocity of the THz light. As demonstrated many times, the plasma can support very strong plasma wave with the high electron density gradient: $n_L / n_0 < 1$ but  $n_L/n_0 \cong 1$ where $n_0 $ is the background electron density. Then, the zeroth order is $n_0(z,t) = n_0 + n_L(z-v_Tt)$. $n_L $  slowly varies compared to the THz light and the laser. 
The computation  of the density perturbation  $\delta n$
 in the presence of $n_L$ due to the ponderomotive interaction between the laser and the THz light is in order. 

%there will be the density perturbation $\delta n$. Our goal is to compute $\delta n$ in the presence of $n_L$. 

 The density response to the ponderomotive potential can be obtained from the continuity equation and the momentum equation: 

\begin{eqnarray}
 \frac{\partial \delta n_e }{\partial t}  &=&-  \mathbf{\nabla} \cdot ( (n_0 + n_L + \delta n_e )\mathbf{v} ) \nonumber \mathrm{,}\\ \nonumber \\
 m_e \frac{d \mathbf{v} }{dt} &=& e\left( \mathbf{\nabla} \phi  - \frac{\mathbf{v}}{c} \times \mathbf{B}\right) \nonumber \mathrm{,} \\ \nonumber
\end{eqnarray}
Combining the above equations with the Poisson equation $\nabla^2 \phi = -4 \pi \delta n_e e $, we can obtain $\delta n$: 
\begin{equation}
\frac{\partial^2 \delta n }{\partial t^2} + \omega_{\mathrm{pe}}(z)^2 \delta n =  - n_0(z,t) \nabla\cdot( \frac{\partial \mathbf{v}}{\partial t} )- \frac{dn_L}{dx}\frac{\partial v}{\partial t}  - \frac{\partial n_L}{\partial t}\frac{d v}{d x}  \mathrm{,} 
\end{equation}
where the velocity $\mathbf{v} $ is generated by the ponderomotive interaction between the THz light and an intense laser.

All physical quantities are expressed as $b(z,t) = b \exp\left(i \omega t - k z\right) + b^*\exp\left(-i \omega t + k z\right) $. Especially $a_{1,T} = eE_{1,T}/m\omega_{1,T} c $ is the laser quiver velocity normalized by the velocity of the light and 
$E_{1,T}$ is the electric field of the laser or the THz light.
Due to the functional form $n_L$, $\partial n_L / \partial t = -v_T dn_L/dx$.   If the velocity $\mathbf{v} $ is caused by a co-propagating laser, 
   $\mathbf{v} = \mathbf{v}( (\omega_1 + \omega_T)t - (k_T + k_1)z)$ by the beating of $a_x a_1$ so that $dv/dx = -c(\omega_x + \omega_1)/ (k_x + k_1) (\partial v/ \partial t)$. 
Also,    $\mathbf{v} = \mathbf{v}((\omega_T - \omega_1)t - ( k_T - k_1)x)$ by the beating of $a_x a_1^*$ so that   $dv/dx = -c(\omega_T - \omega_1)/ (k_T - k_1) (\partial v/ \partial t)$. Putting all these together,  we obtain

\begin{eqnarray}
\frac{\partial^2 \delta n }{\partial t^2} &+&\omega_{\mathrm{pe}}(z)^2 \delta n= 
- n(x) \nabla\cdot( \frac{\partial \mathbf{v}}{\partial t} ) \nonumber 
\nonumber \\ \nonumber 
&-& \frac{dn_L}{dx}\left[1 + A   \right]\frac{\partial v(a_xa_1)}{\partial t}   \\ \nonumber 
&-& \frac{dn_L}{dx}\left[1 + B \right]\frac{\partial v(a_xa_1^*)}{\partial t}  \nonumber  \mathrm{,} \label{eq:den4}
\end{eqnarray}
where $\omega_{\mathrm{pe}}^2(z,t) =4 \pi (n_0+n_L(z,t)) e^2 / m_e$,
$A =  v_T (k_1 + k_T)/(\omega_1 + \omega_T) $ and $B=  v_T (k_1 - k_T)/(\omega_1 - \omega_T) $ for a co-propagating laser and 
$A =  -v_T (k_1 - k_T)/(\omega_1 + \omega_T) $ and $B=  -v_T (k_1 + k_T)/(\omega_1 - \omega_T) $ for counter propagating laser. 
 Then,  the $\delta n$ is given as~\cite{McKinstrie} 

\begin{eqnarray}
\delta n(z)  &=& -   n(z,t)\left( \frac{(ck_1 + ck_T)^2}{ (\omega_1+\omega_T)^2 - \omega_{\mathrm{pe}}^2(z)} \right) a_T a_1 \nonumber \\  
   &+& n(z,t) \left( \frac{(ck_1 - ck_T)^2}{ (\omega_1-\omega_T)^2 - \omega_{\mathrm{pe}}^2(z)} \right) a_T a_1^* \nonumber  \\ 
  &-& \mathbf{i} \frac{dn_L/dz}{k_1 +k_T} \left[1+A\right] 
\left(\frac{(ck_1 + ck_T)^2}{ (\omega_1+\omega_T)^2 - \omega_{\mathrm{pe}}^2(z)}\right) a_T a_1  \nonumber \\ 
    &-& \mathbf{i} \frac{dn_L/dz}{k_1 -k_T} \left[1+B\right] 
\left(\frac{(ck_1 - ck_T)^2}{ (\omega_1-\omega_T)^2 - \omega_{\mathrm{pe}}^2(z)}\right) a_T a_1^* \nonumber \\ \label{eq:minor}
\end{eqnarray} 
In case of the counter-propagating, the $\delta n$ is given as 
\begin{eqnarray}
\delta n(z) &=& -   n(z,t)\left( \frac{(ck_1 - ck_T)^2}{ (\omega_1+\omega_T)^2 - \omega_{\mathrm{pe}}^2(z)} \right) a_T a_1 \nonumber \\  
   &+& n(z,t) \left( \frac{(ck_1 + ck_T)^2}{ (\omega_1-\omega_T)^2 - \omega_{\mathrm{pe}}^2(z)} \right) a_T a_1^* \nonumber  \\ 
  &+& \mathbf{i} \frac{dn_L/dz}{k_1 -k_T} \left[1+A\right] 
\left(\frac{(ck_1 - ck_T)^2}{ (\omega_1+\omega_T)^2 - \omega_{\mathrm{pe}}^2(z)}\right) a_T a_1  \nonumber \\ 
    &+& \mathbf{i} \frac{dn_L/dz}{k_1 +k_T} \left[1+B\right] 
\left(\frac{(ck_1 + ck_T)^2}{ (\omega_1-\omega_T)^2 - \omega_{\mathrm{pe}}^2(z)}\right) a_T a_1^*  \nonumber\mathrm{.} \\ \nonumber \\ \label{eq:major}
\end{eqnarray}
The last two terms in the right side of Eqs.~(\ref{eq:minor}) and (\ref{eq:major}) causes the THz light to decay or  amplify while the first two terms  just modulate the phase of the THz. Assuming $\omega_1 \gg \omega_T$ ($k_1 \gg k_T$) and $v_T \cong c$, and putting the density perturbation into the envelop equation of the THz light~\cite{McKinstrie}, we obtain for the co-propagating laser: 

\begin{equation} 
L_Ta_T = \frac{\omega_{\mathrm{pe}}(z,t)^2}{2\omega_T}
\left( i \beta_1(\eta)  + \beta_2(\eta) \frac{\frac{dn_L}{dz}}{k_1n(z,t)}\right)|a_1|^2 a_T
\label{eq:major2} \mathrm{,} 
\end{equation}
where $\eta = \omega_T/\omega_1$, $\beta_1(\eta) \cong 2$,  and  $\beta_2(\eta) \cong 2/(1-\eta)- 2/(1+\eta)$. If $\eta \ll 1$, this leads to:

\begin{equation} 
L_Ta_T = \frac{\omega_{\mathrm{pe}}(z,t)^2}{\omega_T}
\left( i   + 2\eta \frac{\frac{dn_L}{dz}}{k_1n(z,t)}\right)|a_1|^2 a_T
\label{eq:major5} \mathrm{.} 
\end{equation}
For the counter-propagating laser, 
we obtain:

\begin{equation} 
L_Ta_T = \frac{\omega_{\mathrm{pe}}^2(z,t)}{2\omega_T}
\left( i \beta_1(\eta)  + \beta_2(\eta) \frac{\frac{dn_L}{dz}}{k_1n(z,t)}\right)|a_1|^2 a_T
\label{eq:major3} \mathrm{,}
\end{equation}
where $\beta_1(\eta) \cong (1-\eta)^2/(1+\eta)^2 + (1+\eta)^2/(1-\eta)^2$
and $\beta_2(\eta) \cong - 2\eta \left((1-\eta)/ (1+\eta)^3 +(1+\eta)/(1-\eta)^3\right)$. If $\eta \ll 1$, this leads to: 

\begin{equation} 
L_Ta_T = \frac{\omega_{\mathrm{pe}}^2(z,t)}{\omega_T}
\left( i   - 2\eta \frac{\frac{dn_L}{dz}}{k_1n(z,t)}\right)|a_1|^2 a_T
\label{eq:major4} \mathrm{,}
  \end{equation}
Eqs.~(\ref{eq:major2}) and Eq.~(\ref{eq:major3}) are the major results of this papers. 
For the case of co-propagation (counter-propagating) THz light, 
the THz light will be amplified in the region $dn/dx > 0$ ($dn/dx<0$)

In the resonant Raman scattering between two lasers, 
%the energy is channeled from higher frequency laser to the lower frequency laser. 
the interaction between the lasers excite a Langmuir wave with a phase,  different from the laser ponderomotive potential  by a quarter cycle. 
Due to this phase mismatch, 
the beating current from the Langmuir wave and the  laser quiver  transfers the energy between  the  lasers. 
On the other hand, in a non-resonant case, 
%If the lasers do not satisfy the resonance condition,
the excited density perturbation does not have this phase lag and the energy exchange does not occur. 
%has a phase not suitable for the  energy transfer.   
In the case of the THz and the visible light laser, 
 their  big frequency difference  makes the resonant intercation not feasible. 
% becuase the big difference in their  frequencies make the resonant intercation not feasible. 
As shown here, the first term on the right side of Eqs.~(\ref{eq:major2}) and Eq.~(\ref{eq:major3}) is from  the non-resonant  Raman scattering, which does not decay or amplify the THz light.  However, the density gradient of the plasma could cause this desirable phase mismatch even for the non-resonant excitation, leading to  the decay or amplification of the THz light,  represented as the second terms on the right side of Eqs.~(\ref{eq:major2}) and Eq.~(\ref{eq:major3}).

In many cases, the THz light amplification discussed  has a limited potential, because $\alpha = (dn/dx/k_xn_0) \ll 1$.   Furthermore, as the THz light moves along, $dn/dx$ tends to change the sign and the average of $\alpha$ is even smaller.  
However, if the density gradient is provided by a strong  Langmuir wave supported by plasma and if  the phase velocity of the Langmuir wave is the same  with the group velocity of the THz light, the $\alpha$ can be big and also does not average out in the moving reference frame with the THz light. 
This is the main scenario considered in this paper. 
The case of counter-propagating and co-propagating has almost the same amplification strength. Then, the co-propagation might be more advantageous than the counter-propagation since the interaction time would be much longer. 

As discussed,  
 the key is to make the phase velocity of the Langmuir wave to be the same with the group velocity of the THz light. 
In the case when the forward Raman scattering is utilized, 
the laser has $\omega = \sqrt{\omega_{\mathrm{pe}}^2 + c^2k^2} > ck $ so that the phase velocity $\delta \omega / \delta k > c$. However, it is possible to make 
$(\omega_1 - \omega_2)/|\mathbf{k}_1-\mathbf{k}_2| = v_T < c$ by injecting two lasers in a slightly skewed direction. 
As another possibility, the high frequency microwave from the gyrotron might be also used to excite the Langmuir wave in a larger region~\cite{hidaka, kumar}.

%By utilizing a strong  Langmuir wave supported by plasma and by matching the phase velocity of the Langmuir wave with the group velocity of the THz light, the $\alpha$ can be non-negligible and also does not vary in the moving reference frame with the THz light, which is the main idea of this paper. The case of counter-propagating and co-propagating has almost the same amplification strength. Then, the co-propagation might be more advantageous than the counter-propagation since the interaction time would be much longer. 

As an example, consider a plasma with $n_0 = 10^{17} / / \mathrm{cm}^3$. Consider a far infra-red light with a frequency of 10 THz and the laser with  the wave length of 10 $\mu \mathrm{m}$ so that $\eta = 0.33$.  Let us assume $n_L \cong 0.3  \ n_0$, Then, from Eq.~(\ref{eq:major5}), 
$L_3 a_t  \cong 2 \times 10^{11} \ I_{16} \ / \sec$, where $I_{16}$ is the laser intensity normalized by $10^{16} \ \mathrm{W} / \mathrm{cm}^2 $. 
For a relativistic laser, the gain per length could be as high as 10 (1000) per centimeter if $I_{16} = 1 $ ($I_{16} = 100) $. 
As another example,  consider a plasma with $n_0 = 10^{16} / / \mathrm{cm}^3$. Consider a THz light has a frequency of 3 THz and the laser has  the wave length of 10 \ $\mu \mathrm{m}$. Let us assume $n_L \cong 0.3 \  n_0$, 
$L_3 a_t  \cong 3 \times 10^{10}\  I_{16} \ / \sec$.
 For a relativistic laser, the gain per length could be as high as 1 (100) per centimeter if $I_{16} = 1 $ ($I_{16} = 100) $.
% For a relativistic laser, the gain per length can be as high as 1 per centimeter. 
For a large enough plasma, the THz amplification by many factor is quite possible.  
As illustrated in the above example, as the THz light frequency gets lower, 
the coupling between the laser and THz light is proportional to ($\omega_{\mathrm{pe}}^2 / \omega_1^2$). The scheme proposed will be more efficient and practical in the frequency range between 5 THz to 30 THz rather than between 1 THz and 5 THz.

\bibliography{tera2}% Produces the bibliography via BibTeX.

\end{document}